\documentclass[aps,amsfonts,nofootinbib,superscriptaddress,preprint]{revtex4-1}
\usepackage{comment}
\usepackage[normalem]{ulem} %% Para tachar, comando sout \sout{for the first kinematics}
\makeatletter
\usepackage{amsmath}
\usepackage{amssymb}
\usepackage{appendix}
\usepackage{amsbsy}
\usepackage{bm}
\usepackage{epsfig} 
\usepackage{color}
\usepackage{slashed}
\usepackage{soul}
\usepackage{comment}
\makeatletter
\usepackage{amsmath}
\usepackage{amssymb}
\usepackage{appendix}
\usepackage{amsbsy}
\usepackage{bm}
\usepackage{epsfig} 
\usepackage{color}
\usepackage{slashed}
\newcommand{\stkout}[1]{\ifmmode\text{\sout{\ensuremath{#1}}}\else\sout{#1}\fi}

\newcommand{\bb}{\begin{equation}}

\newcommand{\ee}{\end{equation}}

\newcommand{\eqb}{\begin{eqnarray}}

\newcommand{\eqf}{\end{eqnarray}}

\usepackage[T1]{fontenc}

\begin{document}
\title{  The Laplace Transform of Quantum Gravity}
\author{J. Gamboa}
\email{jorge.gamboa@usach.cl}
\affiliation{Departamento de F\'isica, Universidad de Santiago de Chile, Casilla 307, Santiago, Chile}
\author{F. M\'endez}
\email{fernando.mendez@usach.cl}
\affiliation{Departamento de F\'isica, Universidad de Santiago de Chile, Casilla 307, Santiago, Chile}
\author{Natalia Tapia-Arellano}
\email{ntapiaa@vt.edu}
\affiliation{Center for Neutrino Physics, Department of Physics,
Virginia Tech, Blacksburg, Virginia 24061, USA}
\begin{abstract}
Following analogies with  relativistic point  particles, and Schild strings, we show that the 
Einstein gravity and its  strong coupling regime (or the Planck mass going to 0)
are related to each other through  a Laplace transform. The Feynman propagator of gravity in the strong 
coupling regime satisfies a functional diffusion equation in the three-metric space with the evolution parameter 
being the volume of spacetime. We conjecture that the relationship between both regimes is
consistent with the existence of an evolution operator in which time is replaced by the volume of spacetime..
\end{abstract}
\maketitle

\section{Introduction}
The quantum theory of gravity has been intensively studied in the last seventy years. On the one hand, it is a highly complex 
technical problem and, on the other, a conceptual one that is not yet understood. However, the new ideas incorporated to build 
such a theory will undoubtedly be radically new when it happens. 

Some ideas developed in the study of quantum gravity are, for example \footnote{The literature about this topic 
is extraordinarily  vast, and it is impossible (and surely unpractical) to give a proper bibliographical 
citation to all works on  the subject. We limit ourselves only to original references and reviews. See, for 
example \cite{Carlip:2015asa,Stachel1999}. For a phenomenology approach see \cite{Addazi:2021xuf}.},  
dissipation and  determinism \cite{tHooft:1999imj},  thermodynamics 
\cite{Bekenstein:1972tm,Bekenstein:1973ur,Bekenstein:1974ax},  string theory \cite{Aharony:1999ti}, loop quantum 
gravity  \cite{Ashtekar:2004eh,Rovelli2008}, or  Poincar\'e symmetry deformations of symmetry 
\cite{Lukierski:1991pn,Ballesteros:2017kxj,Addazi:2018jmt}, to mention only a few.

In this paper, we would like to study some technical and conceptual issues of quantum gravity using ideas taken from 
string theory. In particular, Eguchi explored a scheme in \cite{Eguchi:1979qk} for quantizing a string with finite tension 
from  the Schild (tensionless) string \cite{Schild:1976vq}. The idea was applied to extended objects (relativistic membranes) 
\cite{Gamboa:1987tz,Gamboa:1989zd} and it is certainly interesting to study the case of gravity in the strong limit coupling 
\cite{Pilati:1981fy,Henneaux:1981su}. 

The proposed  idea in the present  paper is to establish a series of analogies between particles and strings and then,
when these analogies are understood,  we will apply them to formulate a quantized gravity starting from strong coupling regime \cite{Gamboa:1994aa,Gamboa:1995yn,Gamboa:2000ti}.

The paper is organized as follows; in the next section we discuss an approach to  the quantization of 
particles and strings, where the Laplace transform and the strong limit have a key role. 
In section 3, we apply these ideas   to quantum gravity in the strong coupling regime and we analyze its possible physical implications. Last section is devoted to discussions.

\section{Particles and Schild strings}

In order to present the ideas, let us  start by considering
the Feynman propagator for a relativistic particle of mass $m$  \cite{Feynman:1950ir}
\bb
G[x,x'; m^2] = \int_0^\infty ds ~e^{-\frac{m^2}{2}s} G[x,x'; s], \label{q1}
\ee
where 
\bb
G[x,x'; s]= \int {\cal D} x\, e^{-{\cal S}[x]}, \label{q2}
\ee
and ${\cal S}$ is the Euclidean action of the point  particle.

This idea of introducing mass as a Laplace transform is very useful for studying the 
massless-massive limit transition, and  underlying  to this approach are the proper-time gauge on the one hand, 
and on the other hand, the boundary condition
\bb
x_1^\mu (0)= x_0^\mu, \quad\mbox{and} \quad x_2^\mu (1)= x^\mu_1, \label{qq2}
\ee
which are held fixed upon arbitrary variations. Spacetime has dimension $D$ and therefore
index $\mu=\{0,1,\dots,D-1\}$.

In order to calculate (\ref{q2}) we note that the Euclidean action is  
\bb
{\cal S}= \int_{0}^{1} d\tau \frac{1}{2s} {\dot x}^2, \label{q3}
\ee
which, once replaced in (\ref{q2}), and using (\ref{qq2}) gives 
\bb
G[x,x'; s] = s^{-\frac{D}{2}} ~e^{-\frac{(\Delta x)^2}{2s}}, \label{q4}
\ee
with $\Delta x^\mu = x^\mu_1 - x^\mu_0$. The explicit formula of the Laplace transform (\ref{q1}) is
\bb
G[x,x'; m^2] = \int_0^\infty ds\, s^{-\frac{D}{2}}~e^{-\frac{(\Delta x)^2}{2s}-\frac{m^2}{2}s}, \label{q5}
\ee
the standard integral representation of the Feynman propagator in the proper-time gauge.

The Feynman propagator  satisfies the diffusion equation
\bb
\left(\frac{\partial}{\partial s} - \frac{1}{2} \Box^2 \right) G[x,x'; m^2]= 0.\label{q6}
\ee
From here, the connection to the WKB approach is straightforward. Indeed, applying the operator 
$\frac{\partial}{\partial s} - \frac{1}{2} \Box^2$ to equation (\ref{q2})  we find
\bb
\left(\frac{\partial}{\partial s} - \frac{1}{2} \Box^2 \right) G[x,x'; m^2] = \left[ -\frac{\partial 
{\cal S}}{\partial s}  + \frac{1}{2}\Box^2 S -\frac{1}{2} (\partial {\cal S})^2 \right] G[x,x'; m^2]=0,
\label{diff}
\ee
with $(\partial{\cal S})^2=\partial_\mu{\cal S}\,\partial_\nu{\cal S}\delta^{\mu\nu}$. The Hamilton-Jacobi
equation is obtained from the last equation in  (\ref{diff})
\bb
 -\frac{\partial {\cal S}}{\partial s}= -\frac12 \Box^2 {\cal S} +\frac12 (\partial {\cal S})^2. \label{q7}
\ee
The term $\Box^2{\cal S}$ is a quantum correction and, indeed, in the limit $\hbar\to 0$, this term cancels
and we recover the Hamilton-Jacobi equation of classical mechanics.

Using this result, we turn to the  study of string theory by using the following formal analogy: 
the role of mass corresponds to the tension $T$ in string theory. Therefore, if we can write the Feynman 
propagator as a Laplace transform (as before), then we can naturally make a connection
with the Schild  string  \cite{Schild:1976vq} .

Following the ideas  above, it is natural to think that the equivalent expression  to  (\ref{q2}) is 
\footnote{We will adopt the standard notation $s = N_\perp$ for the rest of the text (see, for example \cite{Nambu:1950rs,Schwinger:1951xk}). Then, the proper-time gauge reads $N_\perp =$ constant.}
\bb 
G[ X(\sigma), X'(\sigma); T] =\int_0^\infty d N_\perp ~e^{- N_\perp~T} ~G[ X(\sigma), X'(\sigma); N_\perp],  \label{q8}
\ee
where 
\bb 
G[ X(\sigma), X'(\sigma); N_\perp] = \int {\cal D} X_\mu (\sigma) ~ e^{-{\cal S}[X(\sigma)]}, \label{q9}
\ee
describes the Schild string dynamics with
\bb 
{\cal S} [X(\sigma)] = \int d^2 \sigma \frac{1}{2 N_\perp} {\dot X}^2 (\sigma). \label{act1}
\ee 
Here the $()^.$ and $()'$ denote $\tau$ and $\sigma$ derivatives, respectively.

These above mentioned expressions,  require careful physical explanation (technical results are explained 
in \cite{Gamboa:1987tz,Gamboa:1989zd} (see also \cite{Ogielski:1980ht})). First, the Schild string is 
the analog to a set of massless particles, and then - if the length of the string is finite as we know - all 
the points of the string move at the speed of light, as implied by the constraint
\bb 
{\cal H}_\perp = \frac{1}{2} P^2 (\sigma) =0. \label{q10}
\ee

However, the string also has to be invariant {under} reparametrizations in the worldsheet. 
That is, the next must be fulfilled 
\bb 
{\cal H}_1 = P_\mu {X'}^\mu =0,  \label{q11}
\ee
implying  that the total Hamiltonian is
\bb 
H = \int d\sigma \left( N_\perp {\cal H}_\perp + N_1 {\cal H}_1 \right) =0, 
\ee 
with $N_\perp,N_1$ two Lagrange multipliers. 
Keeping these results in mind, we set the gauge according to
\bb 
{\dot N}_\perp =0, ~~~~~~~~N_1=0,  \label{q12}
\ee
which is the proper-time gauge \cite{Teitelboim:1981ua}.
In this gauge the most straightforward  choice is setting  $N_{\perp}$ equal to a constant, and that is what we will 
assume. Once this gauge is fixed, the action (\ref{act1}) is obtained.

With this result,  the Laplace transform  turns out to be a consequence of the fact that not only the ends 
of the  string move at the speed of light (as it happens in conventional string theory) \cite{Kalb:1974yc}. Instead, in the 
present case  all the points on the string are moving at the speed of light.

Taking into account these considerations, the calculation of (\ref{q9}) is direct and the result is
\bb 
G[ X(\sigma), X'(\sigma); N_\perp] = s^{-\frac{D}{2}}~e^{-\frac{(\Delta X(\sigma))^2}{2 N_\perp}}. \label{q13}
\ee

The functional diffusion equation is  subtle because the string is an extended {object what makes } necessary 
to add new considerations. 

Indeed, the presence of  two \lq \lq evolution\rq \rq~ parameters, {namely $\sigma$ and $\tau$},
suggests that the true  evolution  parameter (Teichm\"uller one) is the area ($A$) of the string

From the Hamiltonian point of view the equations can be justified as follows; { since the action of the string 
is  $ {\cal S} = \int d \tau d \sigma {\cal L} =\int d A\,{\cal L} $, with  $dA$ an infinitesinal element of 
area of the string, then 
\bb
\frac{d{\cal S}}{dA} = {\cal L}, \label{g16}
\ee
which is true if the integral (the action) depends only on the area of the integration region  
\cite{Moser:1969}. For the Lagrangian under consideration, this property is verified  \cite{Eguchi:1979qk}. }
On the other hand,  $S = S [A, X]$, then
\bb
\frac{d{\cal S}}{dA} = \frac{\partial {\cal S}}{\partial A} +  \frac{\partial {\cal S}}{\partial X}  \frac{\partial X}{\partial A}.  \label{g17}
\ee
If we formally {identify}$ {\dot X} = \frac{\partial X}{\partial A}$ then, by using  (\ref{g16})  and 
the fact ${\cal H} = P{\dot X} -{\cal L}$,  we find
\bb 
{\cal H} \left[ X, \frac{\partial {\cal S}}{\partial X}\right] = - \frac{\partial {\cal S}}{\partial A}, \label{gg16}
\ee
{ where the conjugate momentum has been identified with $\frac{\partial {\cal S}}{\partial X}$}.  
The resulting equation is, therefore, the Hamilton-Jacobi equation for strings with
${\cal H}$ the Hamiltonian density \cite{Hosotani:1985tx,Luscher:1979uq}.

From the quantum point of view -- and as in the case of the relativistic particle discussed above -- the Feynman propagator satisfies the functional diffusion equation \cite{Eguchi:1979qk}
\bb 
\frac{\partial}{\partial A} G[ X(\sigma), X'(\sigma); A]=
\frac{1}{2} \frac{\delta^2}{\delta X^2(\sigma)}  G[ X(\sigma), X'(\sigma); A]. \label{q14}
\ee

Equation (\ref{q9}) and (\ref{q14})  allow us to determine the Hamilton-Jacobi equation
\bb 
\frac{\partial}{\partial A} {\cal S}[X(\sigma)]=  -\frac 12 \frac{\delta^2 {\cal S}[X(\sigma)]}{\delta X^2[\sigma]} + \frac12
\left(\frac{\delta {\cal S}[X(\sigma)]}{\delta X(\sigma)} \right)^2. \label{q15}
\ee

The second derivative, namely $\delta^2 {\cal S} / \delta X^2$, contains the quantum corrections and is 
the main contribution in the systematic treatment of the WKB approximation.  {Naturally, in the present case,
equation (\ref{q15}) is the Schr\"odinger functional equation. In the limit $\hbar\to 0$, the reparametrization
constraint $P_\mu X^\mu \Psi[x(\sigma)] =-\imath\frac{\delta}{\delta X^\mu}  \Psi[x(\sigma)]$ vanishes also,
as it must be due to the gauge choice.}

The study of equations (\ref{q14}) and (\ref{q15}) has been intensively investigated in the past, although the derivation that we have given here is {not the usual one} (for extensions to p-branes see for example
\cite{Gamboa:1987tz}). The solutions in the WKB approximation have been considered in \cite{Luscher:1980fr,Alvarez:1982zi,Migdal:1983qrz,Vassilevich:2003xt}.

\section{Quantum Gravity}
Now we go ahead with the main objective of this  research and we will consider  what happens with quantum gravity.

The Hamiltonian gravity action is 
\bb 
{\cal S} = {\frac1{16\pi G_N}}\int d^4 x \left( \pi_{ij}{\dot g}^{ij} -N_\perp {\cal H}_\perp 
-N_i {\cal H}_i \right), \label{g1}
\ee 
where, as in the case of string theory,  $N_\perp$ and $N_i$ are Lagrange multipliers, $\pi_{ij}$ is canonical momentum and $g_{ij}$ {the metric tensor of the three dimensional space. Latin index, as usual
take values $\{1,2,3\}$ and denote intrinsic quantities of the space section. $G_N$ is the Newton's constant.}

The Hamiltonian constraints are
\eqb 
{\cal H}_\perp &=& 16 \pi G_N~ G_{ijkl} \pi^{ij} \pi^{kl} - \frac{\sqrt{g}}{16 \pi G_N} \left(R^{(3)} - 2 \Lambda\right) \approx 0, \label{g2}
\\
{\cal H}_i&=&  -2 {\pi_i^j}_{;j} \approx 0, \label{g3}
\eqf 
where the supermetric is
\bb 
G_{ijkl}= \frac{1}{2\sqrt{g}} \left(g_{ik} g_{jl} + 
g_{il}g_{jk}-g_{ij}g_{kl} \right), \label{g4}
\ee
{and $\Lambda$ is the cosmological constant.}

The coefficient $16\pi G$ comes from the Einstein-Hilbert action. 
{The Newton's constant $G_N$} has canonical dimension 
{ $-2$ and it can be written as the inverse of the Planck 
mass, that is  $G_N^{-1} = M_{\mbox{\tiny{Pl}}}^2$}. We will use this consideration extensively below.

In order to make more explicit, the analogy between string theory and gravity, let us 
write the Nambu-Goto action
\bb 
{\cal S} = \int d^2 \sigma \left( P_\mu {\dot X}^\mu - N_\perp {\cal H}_\perp - N_1 {\cal H}_1 \right), \label{g5}
\ee 
with 
\eqb 
{\cal H}_\perp &=& \frac{1}{2} \left( \frac{1}{T}P^2 + T {X'}^2 \right) \approx 0, \label{g6}
\\
{\cal H}_1 &=& P_\mu {X'}^\mu \approx 0, \label{g7}
\eqf

If we compare (\ref{g2}) and (\ref{g6}) we see that the analog of a tensionless string corresponds to 
{$M_{\mbox{\tiny{Pl}}}= \frac{1}{G^2} \to 0$} which would be the analog of the massless spectrum in 
string theory. However,  note that even when the analogy is formal, it is technically interesting 
because -- as we will see below -- it is possible to extract technical and physical information 
from an otherwise intractable problem.

The zero Planck mass limit  is also known as the  strong coupling gravity and has been studied in many contexts \cite{Pilati:1981fy,Pilati:1982zi, Teitelboim:1981ua, Gamboa:2000ti,Gamboa:2020yqh}
(for a recent discussion see \cite{Brax:2016vpd} ) and the study of geometric 
properties is part of a very active topic, namely Carroll geometry \cite{Levy1965,Bacry:1968zf}.

Another technical aspect related to Carroll's geometry is that in the limit $M_{\mbox{\tiny{Pl}}} \to 0$, the spacetime is 
causally disconnected { since  the speed of a signal  going from  one point to a neighbor one is zero}  
(that is, $c \to 0$). Parenthetically, this is the reason for the name Carroll 
( {also  note that this limit} is a contraction of the Poincar\'e group). 

Technically {the limit $M_{\mbox{\tiny{Pl}}}\to 0$  is  equivalent to set spatial }
derivatives equal to zero in the  gravity theory.

Keeping these ideas in mind, we can use the proper-time gauge in gravity  \cite{Teitelboim:1981ua} 

\bb 
{\dot N}_\perp =0, ~~~~~~~~N_i=0.  \label{g8}
\ee

Note that the condition ${\dot N}_\perp =0$ establishes that although 
$N_\perp$ does not depend on time, it can depend on spatial coordinates, but without loss of generality, we will assume $N_\perp$ { as  a constant}.

Using this last fact,  the Feynman propagator becomes 
\bb 
G[g_{ij}(2),g_{ij}(1); M_P] = \int_0^\infty dN_\perp~e^{-M_P^2 N_\perp} G[g_{ij}(2),g_{ij}(1); N_\perp], \label{g9}
\ee 
where 
\bb 
G[g_{ij}(2),g_{ij}(1); N_\perp] = \int {\cal D} g_{ij} ~e^{- \int_1^2 d^4x\,\frac{1}{2N_\perp} G_{ijkl} {\dot g}^{ij} {\dot g}^{kl
}},  \label{g10}
\ee 
 
 This equation is the analogue of (\ref{q1}) connecting  the sectors with $M_{\mbox{\tiny{Pl}}} \neq 0 $  with  $ M_{\mbox{\tiny{Pl}}} =0 $. 
 This is  a  very interesting fact because both sectors are presumably  non-perturbatively and, as far as we 
 know,   this has not been previously noticed.
 
{Following the arguments that led us to}  (\ref{g16}), we see that we can find a Hamilton-Jacobi equation 
in a similar way to that of string theory but with a different physical interpretation. Indeed, in gravity the 
action is ${\cal S} = \int d^4 x {\cal L}$, with ${\cal L}$ the Einstein-Hilbert Lagrangian. However,  the 
invariance of reparametrizations {suggests} that we can use the volume of spacetime as a parameter of 
evolution, so {by  analogy with } (\ref{g16}) we have 
 \bb
 \frac{d{\cal S}}{dV} = {\cal L}, 
 \ee
 and the Hamilton-Jacobi equation turn out to be   (\ref{gg16}) with the replacement  $A \to V$.
 
 However (\ref{g10}) implies 
 \bb
 -\frac{\partial {\cal S}}{\partial V} = G^{ijkl} \frac{\delta {\cal S}}{\delta g^{ij}}\frac{\delta {\cal S}}{\delta g^{kl}}  
 -\frac{\delta}{\delta g^{ij}} \left( G^{ijkl} \frac{\delta {\cal S}}{\delta g^{kl}}  \right), \label{g20}
 \ee
 which is, on the one hand, a much more complicated equation and, on the other, it has different physical 
 implications than those we will point out below.

 The first comment is that (\ref{g9}) is a causal relationship between two regions with different scales. 
 However, (\ref{g9}) and the existence of the Laplace transform  are consistent with  a state evolution operator,
 but  with $ V $ playing the role of time in quantum gravity.

  We also emphasize that the Laplace transform (\ref{gg16}) is an unexpected  result because, although it 
  was obtained following the Feynman construction, in the case of gravity,   it turns out to be a relationship 
  that connects  very different energy sectors. 

\section{Discussion}

Following an analogy with string theory (and point relativistic particle), we have written the Feynman 
propagator between two  geometries as the Laplace transform of the propagator describing quantum gravity in the 
strong coupling limit.

Also, like the aforementioned cases, the strong coupling limit propagator satisfies a functional diffusion
equation whit the four dimensional volume playing the role of a time parameter. 

This approach seems to be particularly suitable for treating pure time evolution of the metric, as for 
example the case of cosmology. In fact, for such a case, the action in (\ref{g10}) is proportional to the 
three dimensional volume and the integral contains only the scale factor. To gain an insight on this, consider
the three dimensional metric $g=a^2(t)\mbox{diag}(1,1,1)$, then 
$$
G[a(t_2),a(t_1);N_\perp] =\int {\cal D}a 
\,e^{3V\int_{t_1}^{t_2}  
\frac1{N_\perp}\,\frac{\dot{a}^2}{a^5} dt}.
$$
The action in the exponential turn out to be the action in the strong coupling limit. 
Various issues such as solutions of Einstein’s equations in the strong coupling regime and
explicit quantum formulas have not been written. These and other topics will be the purpose
of a future research work.

\acknowledgments{One of us (J.G.) thanks the Alexander von Humboldt Foundation by support. This research was supported by DICYT 042131GR (J.G.) and 041931MF (F.M.) and by the U.S. Department of Energy under the Award No. DE-SC0020250 (N.T-A.)}

\end{document}